\documentclass[preprint,floats,aps,epsfig,nofootinbib,amssymb,showpacs]{revtex4}
\usepackage{color}
\usepackage{graphicx}
\usepackage{tabularx}
\usepackage{amsmath}
\usepackage{amssymb}
\usepackage{datetime}
\usepackage[left]{lineno}
\usepackage{txfonts}
\usepackage{dcolumn}
\usepackage{bm}

\begin{document}

\title{Direct detection and solar capture of dark matter with momentum and velocity dependent elastic scattering}

\author{Wan-Lei Guo ${}^{1}$}
\email[Email: ]{guowl@itp.ac.cn}

\author{Zheng-Liang Liang ${}^{1,2}$}
\email[Email: ]{liangzl@itp.ac.cn}

\author{Yue-Liang Wu ${}^{1,2}$}
\email[Email: ]{ylwu@itp.ac.cn}

\affiliation{${}^{1}$ State Key Laboratory of Theoretical Physics (SKLTP),\\
Kavli Institute for Theoretical Physics China (KITPC), \\
Institute of Theoretical Physics, Chinese Academy of Science,
Beijing 100190, China \\
${}^{2}$University of Chinese Academy of Sciences, Beijing 100049,
China }

\begin{abstract}

We explore the momentum and velocity dependent elastic scattering
between the dark matter (DM) particles and the nuclei in detectors
and the Sun. In terms of the non-relativistic effective theory, we
phenomenologically discuss ten kinds of momentum and velocity
dependent DM-nucleus interactions and recalculate the corresponding
upper limits on the spin-independent DM-nucleon scattering cross
section from the current direct detection experiments. The DM
solar capture rate is calculated for each interaction. Our
numerical results show that the momentum and velocity dependent
cases can give larger solar capture rate than the usual contact
interaction case for almost the whole parameter space. On the other
hand, we deduce the Super-Kamiokande's constraints on the
solar capture rate for eight typical DM annihilation channels. In
contrast to the usual contact interaction, the Super-Kamiokande and
IceCube experiments can give more stringent limits on the DM-nucleon
elastic scattering cross section than the current direct detection
experiments for several momentum and velocity dependent DM-nucleus
interactions.  In addition, we investigate the mediator
mass's effect on the DM elastic scattering cross section and solar
capture rate.

\end{abstract}


\maketitle

\section{Introduction}

The existence of dark matter (DM) is by now well confirmed
\cite{DM1,DM2}. The recent cosmological observations have helped to
establish the concordance cosmological model where the present
Universe consists of about 68.3\% dark energy, 26.8\% dark matter
and 4.9\% atoms \cite{Ade:2013lta}. Understanding the nature of dark
matter is one of the most challenging problems in particle physics
and cosmology. The DM direct detection experiments may observe the
elastic scattering of DM particles with nuclei in detectors. Current
and future DM direct search experiments may constrain or discover
the DM for its mass $m_D$ and elastic scattering cross section
$\sigma_n$ with nucleon. As well as in the DM direct detection, the
DM particles can also elastically scatter with nuclei in the Sun.
Then they may lose most of their energy and are trapped by the Sun
\cite{DM1}. It is clear that the DM solar capture rate $C_\odot$ is
related to the DM-nucleon elastic scattering cross section
$\sigma_n$. Due to the interactions of the DM annihilation products
in the Sun, only the neutrino can escape from the Sun and reach the
Earth.  Therefore, the water Cherenkov detector Super-Kamiokande
(SK) \cite{SK}, the neutrino telescopes IceCube (IC) \cite{IC1,IC2}
and ANTARES \cite{ANTARES} can also give the information about $m_D$
and $\sigma_n$ through detecting the neutrino induced upgoing muons.

The current experimental results about $\sigma_n$ are based on the
standard DM-nucleus contact interaction which is independent of the
transferred momentum $q$ and the DM velocity $v$. In fact, many DM
scenarios can induce the momentum and velocity dependent DM-nucleus
interactions. For example, the differential scattering cross section
of a long-range interaction will contain a factor $(q^2 +
m_\phi^2)^{-2}$ with $m_\phi$ being the mass of a light mediator $\phi$
\cite{Kumar:2012uh,longrange}. It is worthwhile to stress that the
current experimental results about $\sigma_n$ must be recalculated
for the momentum and velocity dependent DM-nucleus interactions. In
view of this feature, many authors have recently used the momentum
and velocity dependent DM-nucleus interactions to reconcile or
improve the tension between the DAMA annual modulation signal and
other null observations
\cite{longrange,Chang:2009yt,Farina:2011pw,Fitzpatrick:2012ib,Masso:2009mu}.
The new upper limit on $\sigma_n$ can directly affect the maximal
$C_\odot$. On the other hand, we have to recalculate $C_\odot$ for a
fixed $\sigma_n$ when the DM-nucleus interaction is dependent on the
momentum and velocity. For the usual contact interaction, the
current direct search experiment XENON100 \cite{XENON100} provides a
more stringent limit on spin-independent (SI) $\sigma_n$ than the
Super-Kamiokande and IceCube experiments when $m_D \gtrsim 10$ GeV
\cite{SK,IC1,IC2}. We do not know whether this conclusion still
holds for the momentum and velocity dependent DM-nucleus
interactions. It is very necessary for us to systematically
investigate the momentum and velocity dependent DM elastic
scattering in detectors and the Sun.

In this paper, we will explore the momentum and velocity dependent
DM-nucleus interactions and discuss their effects on the SI
$\sigma_n$ and the DM solar capture rate $C_\odot$. New upper limits
on $\sigma_n$ from the XENON100 \cite{XENON100} and XENON10
\cite{XENON10}, and the corresponding maximal $C_\odot$ will be
calculated for these interactions. On the other hand, we shall
deduce the constraints on $C_\odot$ from the latest Super-Kamiokande
results for eight typical DM annihilation channels. In addition, the
mediator mass's effect on $\sigma_n$ and $C_\odot$ will also be
analyzed. This paper is organized as follows: In Sec. II, we outline
the main features of the momentum and velocity dependent DM-nucleus
interactions in direct detection experiments, and derive the
corresponding upper limits on $\sigma_n$. In Sec. III, we
numerically calculate $C_\odot$ for these interactions and give the
general constraints on $C_\odot$ from the Super-Kamiokande and
IceCube. In Sec. IV, we discuss the mediator mass's effect on
$\sigma_n$ and $C_\odot$. Finally, some discussions and conclusions
are given in Sec. V.


\section{Dark matter direct detection}

\subsection{DM event rate \label{2a}}
The event rate $R$ of a DM detector in the direct detection
experiments can be written as
\begin{eqnarray}
R & = & N_T \frac{\rho_{\rm DM}}{m_D}\int \frac{d \sigma_N}{d E_R} d E_R \int _{v_{\rm min}}^{v_{\rm max}} v f(v) d^3 v \;,\nonumber\\
& =& N_T \frac{\rho_{\rm DM}}{m_D} \frac{\pi A^2 m_N
\sigma_n}{\mu_n^2} \int F_N^2(q) d E_R \int_{-1}^{1} d \cos{\theta}
\int_{v_{\rm min}}^{v_{\rm max}} v f(v) F_{\rm DM}^2(q,v) d v \;,
\label{R}
\end{eqnarray}
where $N_T$ is the number of target nucleus in the detector,
$\rho_{\rm DM} = 0.3 {\rm GeV cm}^{-3}$ is the local DM density,
$m_D$ is the DM mass. For the DM-nucleus differential scattering
cross section $d \sigma_N/d E_R$, we have taken the following form
\cite{Farina:2011pw}
\begin{eqnarray}
\frac{d \sigma_N}{d E_R} = A^2 \frac{m_N \sigma_n}{2 v^2 \mu_n^2}
F_N^2(q) F_{\rm DM}^2(q,v)  \;, \label{dsigma}
\end{eqnarray}
where $A$ is the mass number of target nucleus, $\sigma_n$ is the
DM-nucleon scattering cross section. In Eq. (\ref{dsigma}), we have
required that the proton and neutron have the same contribution. The
DM-nucleon reduced mass is given by $\mu_n = m_D m_n/(m_D + m_n)$
where $m_n$ is the nucleon mass. The recoil energy $E_R$ is related
to the transferred momentum $q$ and the target nucleus mass $m_N$
through $q^2 = 2 m_N E_R$. The DM velocity distribution function
$f(v)$ in the galactic frame is usually assumed to be the
Maxwell-Boltzmann distribution with velocity dispersion $v_0 = 220$
km/sec, truncated at the galactic escape velocity $v_{\rm esc} =
544$ km/sec. In the Earth's rest frame, we can derive
\begin{eqnarray}
f(v)=\frac{1}{(\pi v_0^2)^{3/2}} e^{- (\vec{v}+ \vec{v}_e)^2/v_0^2}
\;,
\end{eqnarray}
where $v$ is the DM velocity with respect to the Earth and $v_e
\approx v_\odot = v_0 + 12$km/sec is the Earth's speed relative to
the galactic halo. It is worthwhile to stress that the contribution
of the Earth's orbit velocity to $v_e$ has been neglected since we
do not focus on the annual modulation.  With the help of $|\vec{v}+
\vec{v}_e| \leq v_{\rm esc}$, one can obtain the maximum DM velocity
\begin{eqnarray}
v_{\rm max} = \sqrt{v^2_{\rm esc} - v_e^2 + v_e^2 \cos{\theta}^2}
-v_e \cos{\theta} \;, \label{vmax}
\end{eqnarray}
where $\theta$ is the angle between $\vec{v}$ and  $\vec{v}_e$. For
a given recoil energy $E_R$, one can easily derive the minimum DM
velocity
\begin{eqnarray}
v_{\rm min} = \frac{\sqrt{2 m_N E_R}}{2 \mu_N} \;,
\label{vmin}
\end{eqnarray}
where $\mu_N = m_D m_N/(m_D + m_N)$ is the DM-nucleus reduced mass.
For the nuclear form factor $F_N^2(q)$, we use the Helm form factor
\cite{Helm:1956zz}
\begin{eqnarray}
F_N^2(q) = \left( \frac{3 j_1 (q R_1)}{q R_1} \right)^2 e^{- q^2
s^2}
\end{eqnarray}
with $R_1 = \sqrt{c^2+\frac{7}{3} \pi^2 a^2 - 5 s^2}$ and $c \simeq
1.23 A^{1/3} - 0.60$ fm \cite{Lewin:1995rx}. Here we take $s \simeq
0.9$ fm and $a \simeq 0.52$ fm \cite{Lewin:1995rx}. $j_1(x) =
\sin{x}/x^2 - \cos{x}/x$ is a spherical Bessel function of the first
kind. For the usual contact interaction, the DM form factor $F_{\rm
DM}^2(q,v) = 1$ is independent of the transferred momentum $q$ and
the DM relative velocity $v$. In this paper, we shall focus on some
momentum and velocity dependent DM form factors and discuss their
effects on the DM direct detection cross section  and the DM solar
capture rate.

\subsection{Momentum and velocity dependent DM form factors \label{EO}}

\begin{table}[htb]
\begin{center}
\begin{tabular}{|c|c|c||c|c|c|}
\hline     Case  &   $|{\cal M}|^2$ ($m^2_\phi \gg q^2$)  &  $F_{\rm DM}^2$ ($m^2_\phi \gg q^2$) & Case & $|{\cal M}|^2$ ($m^2_\phi \ll q^2$)  &  $F_{\rm DM}^2$ ($m^2_\phi \ll q^2$) \\
\hline       1   &   $|{\cal M}|^2 \propto 1 $ & $F_{\rm DM}^2 = 1$ &  $ q^{-4}$ & $|{\cal M}|^2 \propto q^{-4} $ & $F_{\rm DM}^2 = q^4_{\rm ref}/q^4$ \\
\hline     $q^2$ &   $|{\cal M}|^2 \propto q^2 $ & $F_{\rm DM}^2 = q^2/q^2_{\rm ref}$ & $q^{-2}$ & $|{\cal M}|^2 \propto q^{-2} $ & $F_{\rm DM}^2 = q^2_{\rm ref}/q^2$\\
\hline     $V^2$ &   $|{\cal M}|^2 \propto V^2 $ & $F_{\rm DM}^2 =V^2/V^2_{\rm ref}$ & $V^2q^{-4}$ & $|{\cal M}|^2 \propto V^2 q^{-4} $ & $F_{\rm DM}^2 =V^2 q^4_{\rm ref}/(V^2_{\rm ref} q^4)$\\
\hline     $q^4$ &   $|{\cal M}|^2 \propto q^4 $ & $F_{\rm DM}^2 = q^4/q^4_{\rm ref}$ & 1 & $|{\cal M}|^2 \propto 1 $ & $F_{\rm DM}^2 = 1 $\\
\hline     $V^4$ &   $|{\cal M}|^2 \propto V^4 $ & $F_{\rm DM}^2 = V^4/V^4_{\rm ref}$ & $V^4q^{-4}$ & $|{\cal M}|^2 \propto V^4 q^{-4} $ & $F_{\rm DM}^2 =V^4 q^4_{\rm ref}/(V^4_{\rm ref} q^4)$\\
\hline  $V^2q^2$ &   $|{\cal M}|^2 \propto V^2 q^2 $ & $F_{\rm DM}^2 = V^2 q^2/(V^2_{\rm ref} q^2_{\rm ref}) $ &  $V^2q^{-2}$  & $|{\cal M}|^2 \propto V^2 q^{-2} $ & $F_{\rm DM}^2 =V^2 q^2_{\rm ref}/(V^2_{\rm ref} q^2)$\\
\hline
\end{tabular}
\end{center}
\vspace{-0.2cm} \caption{The momentum and velocity dependent DM form
factors for the heavy and light mediator mass scenarios with $V^2 =
v^2 - q^2/(4 \mu_N^2)$, $q_{\rm ref} = 100$ MeV and $V_{\rm ref} =
v_0$. } \label{FDM}
\end{table}

Usually, one can build some DM models and exactly calculate the DM
direct detection cross section. On the other hand, the DM-nucleus
interaction can be generally constructed from 16 model-independent
operators in the non-relativistic (NR) limit
\cite{Fan:2010gt,Fitzpatrick:2012ix}. Any other scalar operators
involving at least one of the two spins can be expressed as a linear
combination of the 16 independent operators with SI coefficients
that may depend on ${q}^2$ and $\vec{V}^2\equiv
(\vec{v}-\vec{q}/(2\mu_N))^2 = v^2-q^2/(4 \mu_N^2)$. It is
convenient for us to phenomenologically analyze the momentum and
velocity dependent DM-nucleus interactions from these NR operators.
Here we only focus on the following four SI NR operators in the
momentum space \cite{Fan:2010gt,Fitzpatrick:2012ix}:
\begin{eqnarray}
{\cal{O}}_1 & = & 1 \;, \nonumber\\
{\cal{O}}_2 & = & i \vec{s}_D \cdot \vec{q}  \;,  \nonumber\\
{\cal{O}}_3 & = & \vec{s}_D \cdot \vec{V} \;,  \nonumber\\
{\cal{O}}_4 & = & i \vec{s}_D \cdot (\vec{V} \times \vec{q}) \;.
 \label{operator}
\end{eqnarray}
Considering the possible contributions of $q^2$ or $V^2$ in the
coefficients, we phenomenologically discuss five kinds of momentum and
velocity dependent DM form factors $F_{\rm DM}^2(q,v)$ up to $q$ and
$\vec{V}$ quartic terms in the amplitude squared $|{\cal M}|^2$. The
five DM form factors and the usual contact interaction case have been
listed in the third column of Table \ref{FDM}. Since the transferred
momentum $q$ in many direct detection experiments is order of 100
MeV, we take $q_{\rm ref} = 100$ MeV as the reference transferred
momentum to normalize $q$. Similarly, we use $V_{\rm ref} = v_0$ to
normalize $V = \sqrt{v^2 - \frac{q^2}{4 \mu_N^2}}$. Here we have
assumed that the mass $m_\phi$ of mediator between DM particles and
quarks is far larger than the transferred momentum $q$, namely
$m^2_\phi \gg q^2$. If $m^2_\phi \ll q^2$, $F_{\rm DM}^2(q,v)$
should contain the factor $1/q^4$ which comes from the squared
propagator $(q^2 + m_\phi^2)^{-2}$. For the light mediator mass
scenario, the corresponding 6 kinds of $F_{\rm DM}^2(q,v)$ cases
have been listed in the sixth column of Table \ref{FDM}. In Sec. IV,
we shall discuss the $m^2_\phi \sim {\cal{O}}(q^2)$ scenario through
varying $m_\phi$.

\subsection{New upper limits on $\sigma_n$}

In this paper, we do not try to reconcile the tension between the
DAMA annual modulation signal and other direct detection exclusions
by use of the momentum and velocity dependent DM-nucleus
interactions. Here we only focus on the null observations and the
corresponding upper limits on $\sigma_n$ which are relevant to the
maximal DM solar capture rate. Currently, the most stringent limit
on $\sigma_n$ comes from XENON100 \cite{XENON100} and XENON10
\cite{XENON10}. It should be mentioned that this limit is only valid
for the usual contact interaction, namely $F_{\rm DM}^2(q,v)= 1$.
For the momentum and velocity dependent $F_{\rm DM}^2(q,v)$, we
should recalculate their limits from the reported results of
XENON100 and XENON10.

The recoil energy window of the DM search region in the XENON100 is
chosen between $3\sim 30$ photoelectrons (PE), corresponding to $6.6 \;
{\rm keV} \leq E_R \leq 43.3$ keV. The relation of $E_R$ and PE
number $S1$ is given by \cite{XENON100}
\begin{eqnarray}
S1(E_R) = 3.73 \; {\rm PE} \times E_R \times {\cal L}_{\rm eff}\;,
\label{S1}
\end{eqnarray}
where ${\cal L}_{\rm eff}$ is the scintillation efficiency which has
been measured above 3 keV. The ${\cal L}_{\rm eff}$ parametrization
can be found in Ref. \cite{Aprile:2011hi}. Here we assume that the
produced PE number of a nucleus recoil event satisfies the
Poissonian distribution and Eq. (\ref{S1}) denotes the mean value.
In this case, the event with $E_R < 6.6$ keV will have a
non-vanishing probability to generate a $S1$ signal above 3 PE. For
the new lower threshold of $E_R$, we take $E_R \geq 3.0$ keV which
can pass the ionization yield $S2$ cut
\cite{Savage:2010tg,Farina:2011pw}.

The search recoil energy range of XENON10 is $1.4 \; {\rm keV} \leq
E_R \leq 10.0$ keV \cite{XENON10}. For $4 \; {\rm GeV}  \lesssim m_D
\leq 20$ GeV, one can always find some parameter space among $1.4 \;
{\rm keV} \leq E_R \leq 10.0$ keV to satisfy  $v_{\rm min} < v_{\rm
max}$. Therefore, we directly input  $1.4 \; {\rm keV} \leq E_R \leq
10.0$ keV into Eq. (\ref{R}) for the XENON10 analysis. Note that the
upper limit with $v_0 =230$ km/sec and $v_{\rm esc} =600$ km/sec in
Ref. \cite{XENON10} has been replaced by the corresponding limit
with $v_0 =220$ km/sec and $v_{\rm esc} =544$ km/sec in the
following parts.

\begin{figure}[htb]
\begin{center}
\includegraphics[scale=0.42]{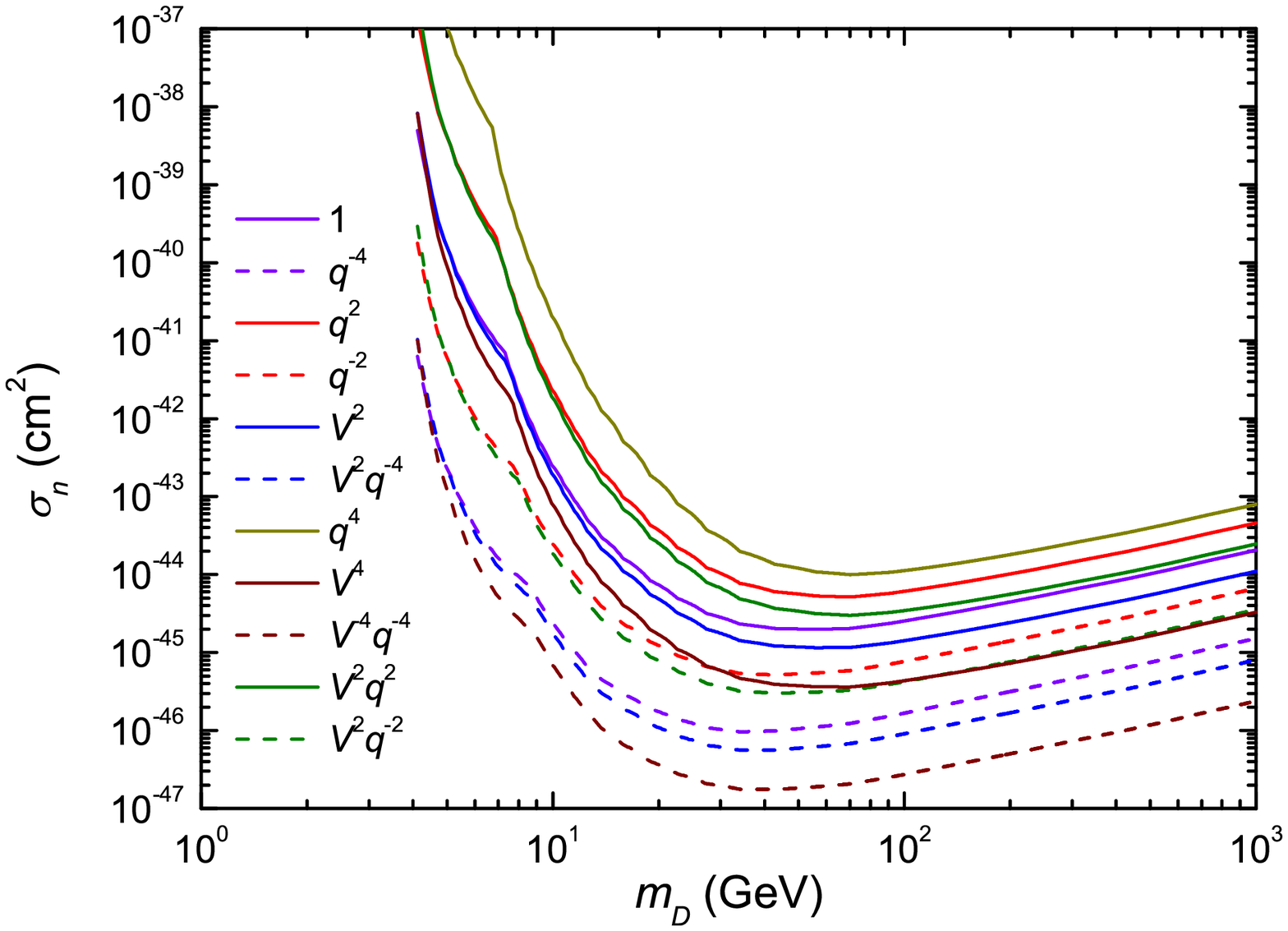}
\includegraphics[scale=0.42]{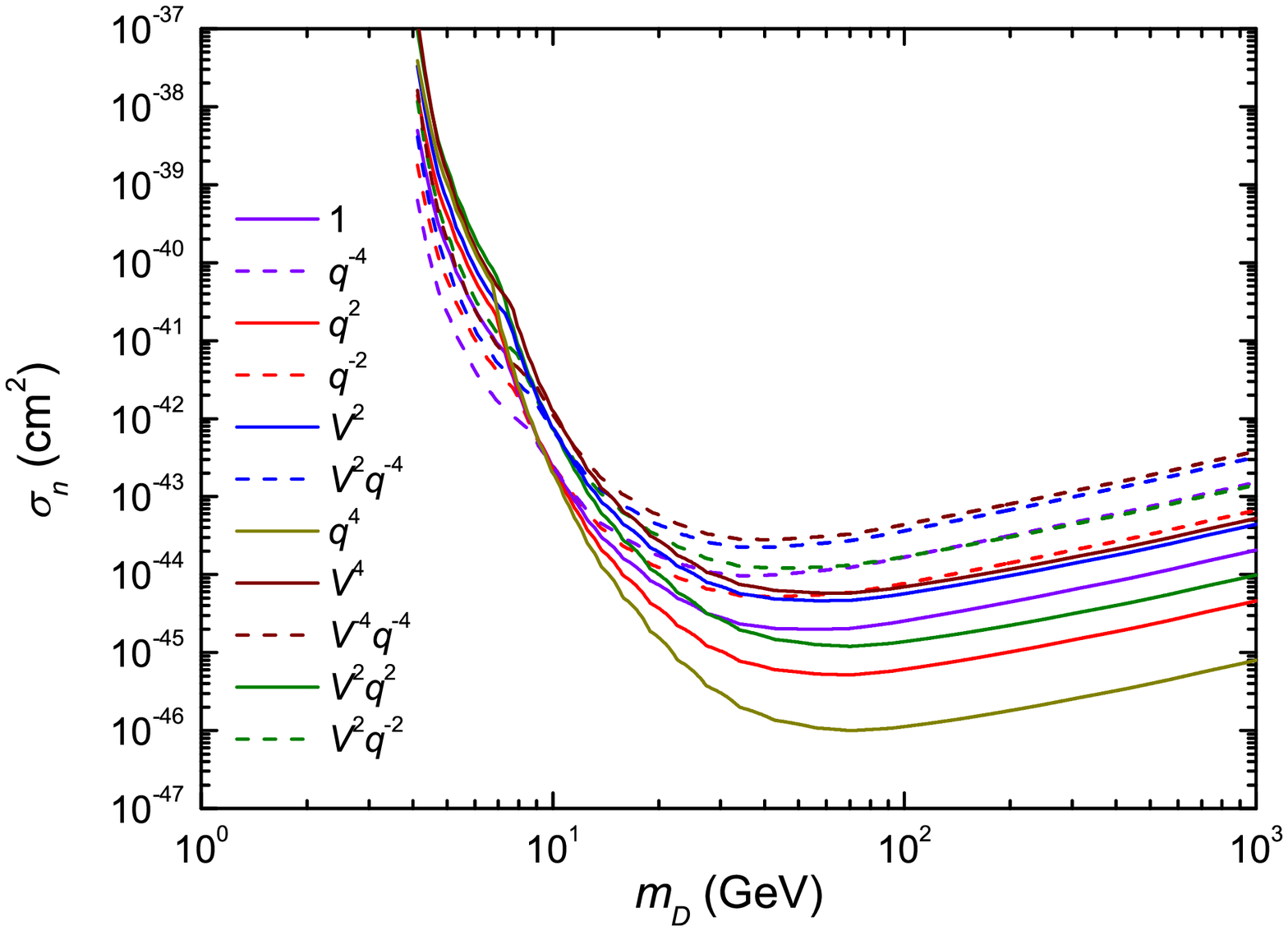}
\end{center}
\vspace{-0.5cm}\caption{ The new upper limits on $\sigma_n$ for
different $F_{\rm DM}^2(q,v)$ with $q_{\rm ref} = 100$ MeV; $V_{\rm
ref} = v_0$ (left panel) and $q_{\rm ref} = 10\sqrt{10}$ MeV;
$V_{\rm ref} = 2 v_0$ (right panel) from the XENON100 and XENON10.}
\label{DI}
\end{figure}

Requiring the same event rate $R$ for different $F_{\rm DM}^2(q,v)$,
we deduce new bounds about $\sigma_n$ for each $F_{\rm DM}^2(q,v)$
from the $F_{\rm DM}^2(q,v)=1$ case (the reported limits of XENON100
and XENON10). Our numerical results have been shown in the left
panel of Fig. \ref{DI}.  The same color solid and dashed lines
describe the heavy and the corresponding light mediator mass
scenarios, respectively. In Fig. \ref{DI}, the number 1 denotes the $F_{\rm
DM}^2(q,v)=1$ case, $V^2 q^{-2}$ denotes the $F_{\rm DM}^2(q,v)= V^2
q^2_{\rm ref}/ (V^2_{\rm ref} q^2)$ case, and so on. It is
meaningless for us to compare different lines since these limits are
dependent on $q_{\rm ref}$ and $V_{\rm ref}$. For illustration, we
plot our numerical results with $q_{\rm ref} = 10\sqrt{10}$ MeV and
$V_{\rm ref} = 2 v_0$ in the right panel of Fig. \ref{DI}. Some
kinks around $m_D = 8$ GeV arise from the different slopes of the
predicted limits of XENON100 and XENON10. It is should be mentioned
that the new upper bound from the XENON100 and XENON10 is still the
most stringent limit for each $F_{\rm DM}^2(q,v)$ when we
recalculate other experimental results \cite{CDMS}.

\section{Dark matter solar capture}

When the halo DM particles elastically scatter with nuclei in the
Sun, they may lose most of their energy and are trapped by the Sun
\cite{DM1}. On the other hand, the DM annihilation in the Sun
depletes the DM population. The evolution of the DM number $N$ in
the Sun is given by the following equation \cite{Griest:1986yu}:
\begin{eqnarray}
\dot{N} = C_\odot  - C_A N^2 \;, \label{N}
\end{eqnarray}
where the dot denotes differentiation with respect to time. The DM
solar capture rate $C_\odot$ in Eq. (\ref{N}) is proportional to the
DM-nucleon scattering cross section $\sigma_n$. In the next
subsection, we shall give the exact formulas to calculate $C_\odot$.
The last term $C_A N^2$ in Eq. (\ref{N}) controls the DM
annihilation rate in the Sun. The coefficient $C_A$ depends on the
thermal-average of the annihilation cross section times the relative
velocity $\langle \sigma v \rangle$ and the DM distribution in the
Sun. To a good approximation, one can obtain $C_A = \langle \sigma v
\rangle/V_{\rm eff}$, where $V_{\rm eff} = 5.8 \times 10^{30} \;
{\rm cm^3} ( 1 {\rm GeV}/{m_D} )^{3/2}$ is the effective volume of
the core of the Sun \cite{Griest:1986yu,Gould:1987ir}. In Eq.
(\ref{N}), we have neglected the evaporation effect since this
effect is very small when $m_D \gtrsim 4$ GeV
\cite{Gould:1987ju,Hooper:2008cf}. One can easily solve the
evolution equation and derive the DM solar annihilation rate
\cite{Griest:1986yu}
\begin{eqnarray}
\Gamma_A = \frac{1}{2} C_\odot \tanh^2 (t_\odot \sqrt{ C_\odot C_A}
) \;, \label{GammaA}
\end{eqnarray}
where $t_\odot \simeq 4.5$ Gyr is the age of the solar system. If
$t_\odot \sqrt{ C_\odot C_A} \gg 1$, the DM annihilation rate
reaches equilibrium with the DM capture rate. In this case, we
derive the maximal DM annihilation rate $\Gamma_A = C_\odot/2$. It
is clear that the DM annihilation signals from the Sun are entirely
determined by $C_\odot$.

\subsection{DM solar capture rate and annihilation rate}

By use of the DM angular momentum conservation in the solar
gravitational field, one can obtain the following DM capture rate
$C_{\odot}$ \cite{Gould:1987ir}:
\begin{eqnarray}
C_{\odot} = \sum_{N_i} \int 4 \pi r^2 d r \int \frac{f(u)}{u} \omega
\Omega_{N_i}(\omega) d^3 u  \label{csun}
\end{eqnarray}
with
\begin{eqnarray}
f(u)=\frac{1}{(\pi v_0^2)^{3/2}} e^{- (\vec{u}+
\vec{v}_\odot)^2/v_0^2} \;, \label{fu}
\end{eqnarray}
where $f(u)$ is the DM velocity distribution, $u$ is the DM velocity
at infinity with respect to the Sun's rest frame, $v_\odot = v_0
+12$ km/sec is the Sun's speed relative to the galactic halo.
$\Omega_{N_i}(\omega)$ is the rate per unit time at which a DM
particle with the incident velocity $\omega$ scatters to an orbit
within the Jupiter's orbit. $\Omega_{N_i}(\omega)$ is given by
\begin{eqnarray}
\Omega_{N_i}(\omega) = n_{N_i}(r) \sigma_{N_i}(\omega) \omega
\rho_{\rm DM}/ m_D \;, \label{Omega}
\end{eqnarray}
where $n_{N_i}(r)$ and $\omega(r)= \sqrt{u^2 + v^2_{\rm esc}(r)}$
are the number density of element ${N_i}$ and the DM incident
velocity at radius $r$ inside the Sun, respectively. The escape
velocity $v_{\rm esc}(r)$ from the Sun at the radius $r$ can be
approximately written as $v^2_{\rm esc}(r) = v^2_c - (v_c^2 - v_s^2)
M(r)/M_\odot$ \cite{Gould:1991hx}, where $v_c =1354$ km/sec and $v_s
= 795$ km/sec are the escape velocity at the Sun's center and
surface, respectively. $M_\odot = 1.989 \times 10^{33} {\rm g}$ is
the solar mass and $M(r) $ is the mass within the radius $r$.
$\sigma_{N_i}(\omega)$ in Eq. (\ref{Omega}) is the scattering cross
section between a stationary target nucleus ${N_i}$ in the Sun and
an incident DM particle with velocity $\omega$. The non-relativistic
effective theory allows us to express $\sigma_{N_i}(\omega)$ as
\begin{eqnarray}
\sigma_{N_i}(\omega) = \frac{A_i^2 \sigma_n}{2 \omega^2 \mu_n^2}
\int_{q_{\rm min}}^{2 \mu_{N_i} \omega} F_{N_i}^2(q) F_{\rm DM}^2
(q, \omega) q dq \;, \label{sigmaA}
\end{eqnarray}
where $q_{\rm min} = \sqrt{m_D m_{N_i} [u^2 + v_{\rm esc}^2(r=5.2
{\rm AU})]}$ is the minimum transferred momentum needed for capture
and $v_{\rm esc}(r=5.2 {\rm AU}) = 18.5$ km/s denotes the DM escape
velocity from the Sun at the Jupiter's orbit
\cite{Kumar:2012uh,Peter:2009mk}.

\begin{figure}[htb]
\begin{center}
\includegraphics[scale=0.42]{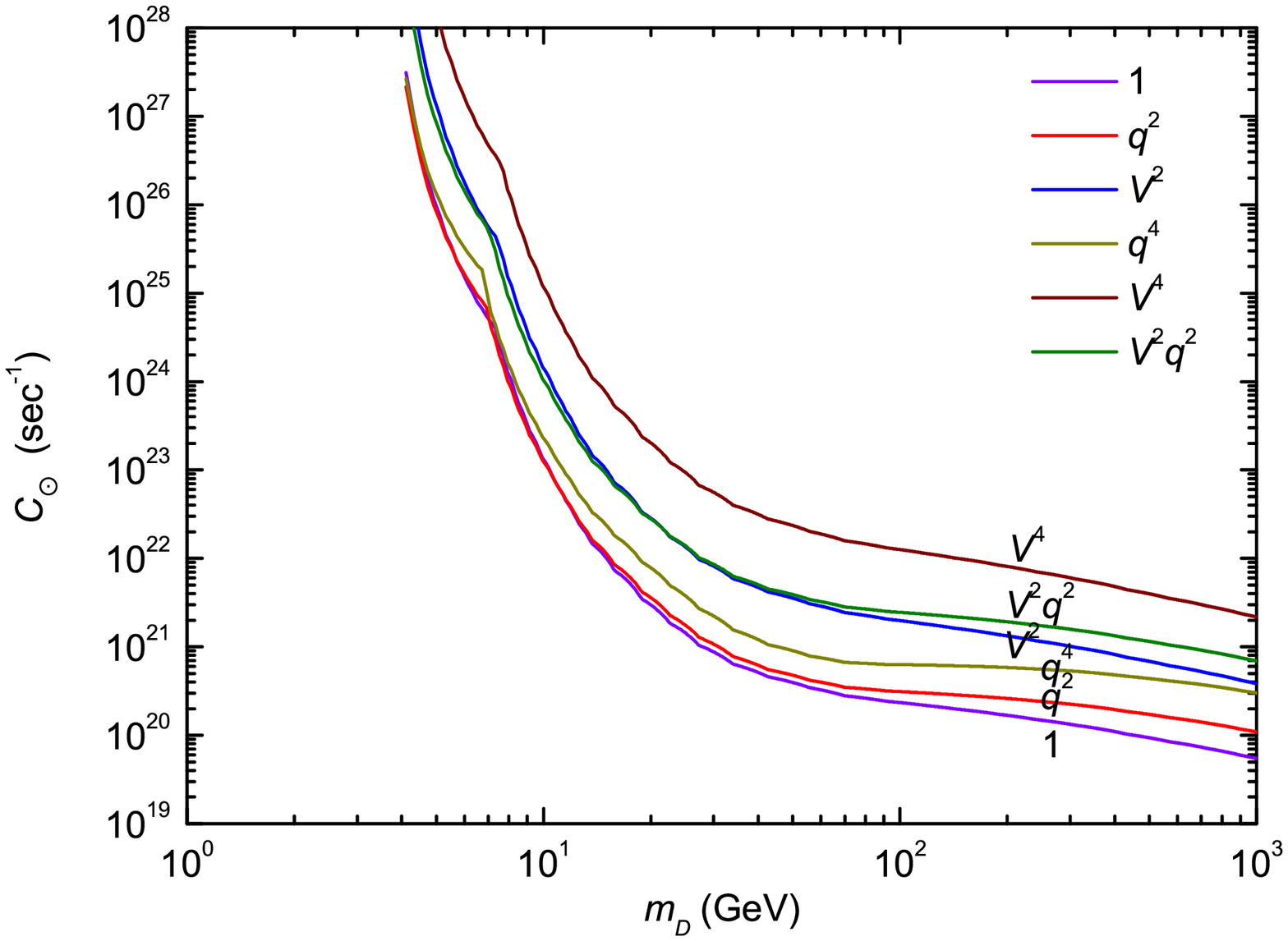}
\includegraphics[scale=0.42]{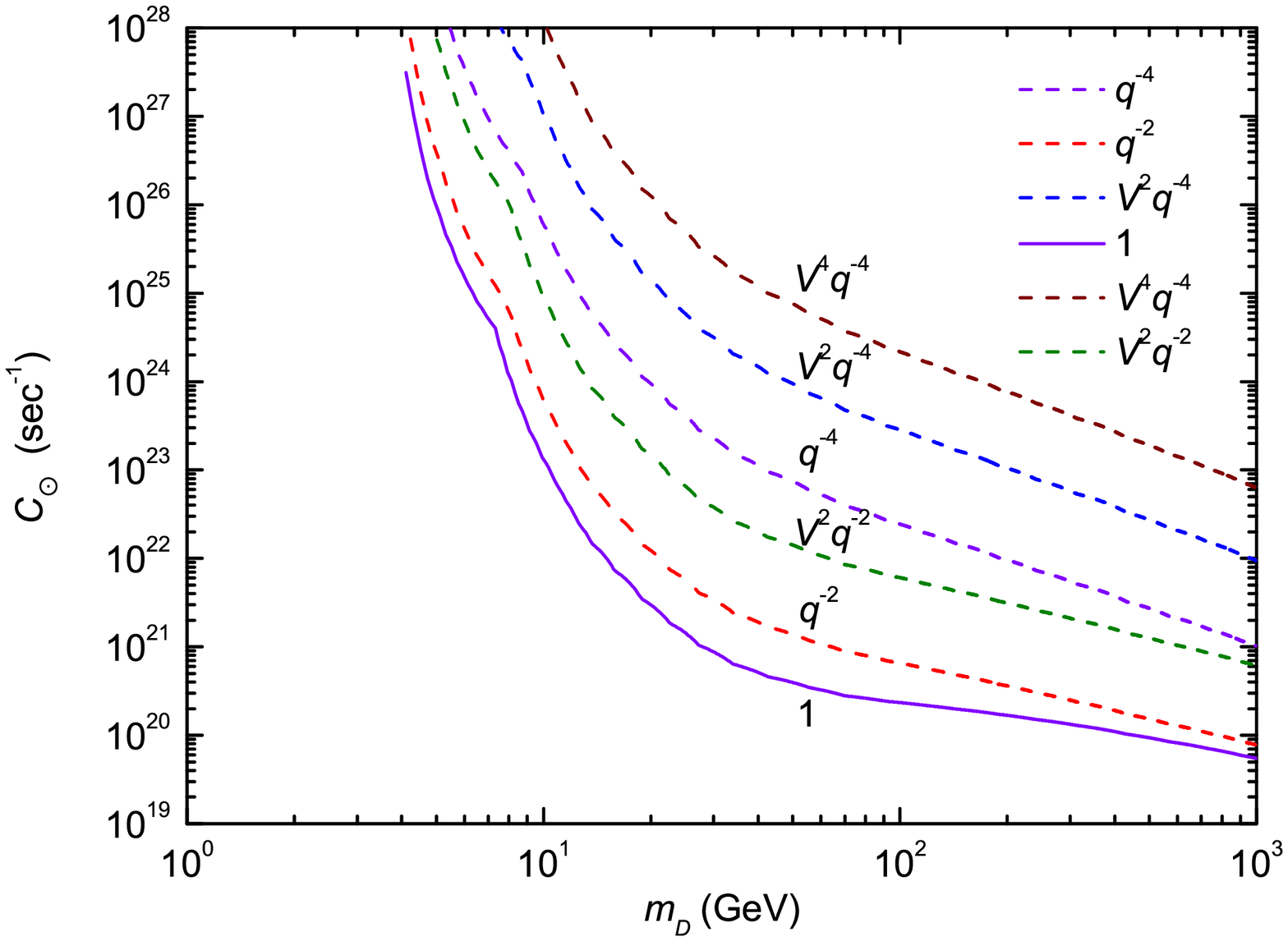}
\end{center}
\vspace{-0.5cm}\caption{ The predicted maximal DM solar capture
rates $C_\odot$ for the heavy (left panel) and light (right panel)
mediator mass scenarios.} \label{CsunHL}
\end{figure}

In our calculation, we sum over the following elements in the Sun:
$^1$H, $^4$He, $^{12}$C, $^{14}$N, $^{16}$O, $^{17}$O, Ne, Mg, Si, S
and Fe. The number densities $n_{N_i}(r)$ of these elements and
$M(r)$ can be obtained from the calculation of the standard solar
model (SSM). Here we employ the SSM GS98 \cite{Serenelli:2009yc} to
calculate the DM solar capture rate $C_\odot$ in Eq. (\ref{csun})
with the help of $\sigma_n$ in the left panel of Fig. \ref{DI}. Our
numerical results have been shown in Fig. \ref{CsunHL}. We find that
for almost whole of the $m_D$ parameter space
the predicted $C_\odot$ from the standard contact interaction is
smaller than those from the momentum and velocity dependent DM form
factor cases. This means that the momentum and velocity dependent DM form factor cases
can give larger DM annihilation signals than that from the usual
contact interaction case. The same color solid and dashed lines in
Fig. \ref{CsunHL} describe the heavy and the corresponding light
mediator mass scenarios, respectively. The light mediator mass
scenario can usually produce the larger $C_\odot$ than the
corresponding heavy mediator mass scenario. However one can derive
the opposite conclusion for the $q^4$ and $1$ cases. It should be
mentioned that our numerical results in Fig. \ref{CsunHL} are
independent of $q_{\rm ref}$ and $V_{\rm ref}$.

\subsection{Constraints from the Super-Kamiokande and IceCube}

Due to the interactions of the DM annihilation products in the Sun,
only the neutrino can escape from the Sun and reach the Earth.  For
the given DM mass and DM annihilation channel $\alpha$, the
differential muon neutrino  flux at the surface of the Earth from
per DM pair annihilation in the Sun can be written as
\begin{eqnarray}
\frac{d \Phi^{\alpha}_{\nu_{\mu}} }{d E_{\nu_{\mu}} } =
\frac{\Gamma_A}{4 \pi R_{\rm ES}^2} \frac{d N^{\alpha}_{\nu_{\mu}}
}{d E_{\nu_{\mu}} } \;, \label{dphide}
\end{eqnarray}
where $R_{\rm ES} = 1.496 \times 10^{13}$ cm is the Earth-Sun
distance. $d N^{\alpha}_{\nu_\mu}/d E_{\nu_\mu}$ denotes the energy
distribution of neutrinos at the surface of the Earth produced by
the final state $\alpha$ through hadronization and decay processes
in the core of the Sun. It should be mentioned that some produced
particles, such as the muon and abundantly produced light hadrons
can lose almost total energy before they decay due to their
interactions in the Sun. In addition, we should consider the
neutrino interactions in the Sun and neutrino oscillations. In this
paper, we use the program package WimpSim \cite{WimpSim} to
calculate $d N^{\alpha}_{\nu_\mu}/d E_{\nu_\mu}$ with the following
neutrino oscillation parameters \cite{Tortola:2012te,An:2012eh}:
\begin{eqnarray}
\sin^2 \theta_{12} = 0.32 ,\;\; \sin^2 \theta_{23} = 0.49, & &
\sin^2 \theta_{13} = 0.026, \;\; \delta=0.83 \pi, \nonumber \\
\Delta m_{21}^2 = 7.62 \times 10^{-5} {\rm eV}^2,  & & \Delta
m_{31}^2 = 2.53 \times 10^{-3} {\rm eV}^2\;.
\end{eqnarray}
In addition, we should also calculate the differential muon
anti-neutrino flux which can be evaluated by an equation similar to
Eq. (\ref{dphide}).

These high energy neutrinos interact with the Earth rock or ice to
produce upgoing muons which may be detected by the water Cherenkov
detector Super-Kamiokande \cite{SK} and the neutrino telescope
IceCube \cite{IC1,IC2}. Due to the produced muons scattered from the
primary neutrino direction and the multiple Coulomb scattering of
muons on route to the detector, the final directions of muons are
spread. For $10 \;{\rm GeV} \leq m_D \leq 1000$ GeV, the cone
half-angle which contains more than $90\%$ of the expected event
numbers ranges from $6^\circ$ to $30^\circ$ for the Super-Kamiokande
when we assume the $b \bar{b}$ annihilation channel. The cone
half-angles will be smaller for the other DM annihilation channels
considered in this paper with the same DM mass. In terms of the
results of cone half-angle $\theta$ in Tables 1 and 2 of Ref.
\cite{SK}, we conservatively take some reasonable $\theta$ for other
DM annihilation channels and several representative DM masses as
shown in Table \ref{N90phi}.

\begin{table}[htb]
\footnotesize
\begin{center}
\begin{tabular}{|c|c|c|c|c|c|c||c|c|c|c|c|c|c|}
\hline   {\scriptsize Channel}  &  $m_D$  &  $\theta$  &   $F^i \; (\%)$  & $N_{90}$ & $\phi_{\mu}$ & $\Gamma_A$ (sec$^{-1}$) & {\scriptsize Channel}  &  $m_D$  &  $\theta$  &   $F^i \; (\%)$  & $N_{90}$ & $\phi_{\mu}$ & $\Gamma_A$(sec$^{-1}$) \\
\hline
\hline  $\nu_e \bar{\nu}_e$ &  4 & 30$^\circ$ & 93.1; 5.5; 1.4 & 15.65 & 9.4 & $7.2 \times 10^{24}$ & $\nu_\mu \bar{\nu}_\mu$ &  4 & 30$^\circ$ & 93.1; 5.5; 1.4 & 15.65 & 9.4 & $6.7 \times 10^{24}$ \\
\hline  $\nu_e \bar{\nu}_e$ &  6 & 30$^\circ$ & 87.0; 9.9; 3.1 & 16.62 & 10.0 &  $2.7 \times 10^{24}$ & $\nu_\mu \bar{\nu}_\mu$ &  6 & 30$^\circ$ & 87.0; 9.9; 3.1 & 16.62 & 10.0  & $2.4 \times 10^{24}$ \\
\hline  $\nu_e \bar{\nu}_e$ &  10 & 30$^\circ$ & 73.4; 19.3; 7.3 & 19.14 & 11.5 &  $8.3 \times 10^{23}$ & $\nu_\mu \bar{\nu}_\mu$ &  10 & 30$^\circ$ & 73.4; 19.3; 7.3 & 19.14 & 11.5  & $6.7 \times 10^{23}$ \\
\hline  $\nu_e \bar{\nu}_e$ &  $10^2$ & 7$^\circ$ & 15.3; 58.6; 26.1 & 7.33 & 4.4 &  $6.3 \times 10^{21}$ & $\nu_\mu \bar{\nu}_\mu$ &  $10^2$ & 7$^\circ$ & 17.8; 56.9; 25.3 & 7.35 & 4.4  & $1.9 \times 10^{21}$ \\
\hline  $\nu_e \bar{\nu}_e$ &  $10^3$ & 3$^\circ$ & 14.4; 53.6; 32.0 & 4.64 & 2.8 &  $1.5 \times 10^{21}$ & $\nu_\mu \bar{\nu}_\mu$ &  $10^3$ & 3$^\circ$ & 17.4; 52.2; 30.4 & 4.60& 2.8  & $9.6 \times 10^{20}$ \\
\hline
\hline  $\nu_\tau \bar{\nu}_\tau$ &  4 & 30$^\circ$ & 93.1; 5.5; 1.4 & 15.65 & 9.4 &  $6.7 \times 10^{24}$ & $\tau^+ \tau^-$ &  4 & 30$^\circ$ & 96.1; 3.2; 0.7 & 15.22 & 9.1  & $1.1 \times 10^{26}$ \\
\hline  $\nu_\tau \bar{\nu}_\tau$ &  6 & 30$^\circ$ & 87.0; 9.9; 3.1 & 16.62 & 10.0 &  $2.4 \times 10^{24}$ & $\tau^+ \tau^-$ &  6 & 30$^\circ$ & 94.9; 4.1; 1.0 & 15.40 & 9.2  & $2.0 \times 10^{25}$ \\
\hline  $\nu_\tau \bar{\nu}_\tau$ &  10 & 30$^\circ$ & 73.4; 19.3; 7.3 & 19.14 & 11.5 &  $6.7 \times 10^{23}$ & $\tau^+ \tau^-$ &  10 & 30$^\circ$ & 91.3; 6.7; 2.0 & 15.94 & 9.5  & $4.4 \times 10^{24}$ \\
\hline  $\nu_\tau \bar{\nu}_\tau$ &  $10^2$ & 7$^\circ$ & 20.8; 54.9; 24.3 & 7.35 & 4.4 &  $3.0 \times 10^{21}$ & $\tau^+ \tau^-$ &  $10^2$ & 7$^\circ$ & 44.8; 39.2; 16.0 & 6.81 & 4.1  & $1.4 \times 10^{22}$ \\
\hline  $\nu_\tau \bar{\nu}_\tau$ &  $10^3$ & 3$^\circ$ & 28.6; 48.2; 23.2 & 4.42 & 2.6 &  $4.9 \times 10^{20}$ & $\tau^+ \tau^-$ &  $10^3$ & 3$^\circ$ & 27.9; 48.8; 23.3 & 4.43 & 2.7  & $5.8 \times 10^{20}$ \\
\hline
\hline  $W^+ W^-$ &  81 & 8$^\circ$ & 44.6; 39.4; 16.0 & 8.38 & 5.0 &  $6.2 \times 10^{22}$ & $ Z Z$ &  92 & 8$^\circ$ & 47.4; 37.4; 15.2 & 8.20 & 4.9  & $4.0 \times 10^{22}$ \\
\hline  $W^+ W^-$ &  $10^2$ & 7$^\circ$ & 43.4; 40.1; 16.5 & 6.86 & 4.1 &  $3.3 \times 10^{22}$ & $ Z Z$ &  $10^2$ & 7$^\circ$ & 46.6; 37.9; 15.5 & 6.74 & 4.0  & $2.8 \times 10^{22}$ \\
\hline  $W^+ W^-$ &  $10^3$ & 3$^\circ$ & 34.4; 44.6; 21.0 & 4.31 & 2.6 &  $1.9 \times 10^{21}$ & $ Z Z$ &  $10^3$ & 3$^\circ$ & 40.6; 40.6; 18.8 & 4.18 & 2.5  & $1.7 \times 10^{21}$ \\
\hline
\hline  $b \bar{b}$ &  6 & 30$^\circ$ & 96.7; 2.7; 0.6 & 15.14 & 9.1 &  $1.5 \times 10^{27}$ &   &    &   &   & &  & \\
\hline  $b \bar{b}$ &  10 & 30$^\circ$ & 95.6; 3.6; 0.8 & 15.29 & 9.2 &   $1.3 \times 10^{26}$ &  &    &   &   & &  & \\
\hline  $b \bar{b}$ &  $10^2$ & 10$^\circ$ & 77.2; 16.6; 6.2 & 10.93 & 6.5 &  $5.9 \times 10^{23}$ & $ t \bar{t}$  & 175   & 10$^\circ$  & 61.2; 27.7; 11.1  & 12.26 & 7.3   & $4.4 \times 10^{22}$ \\
\hline  $b \bar{b}$ &  $10^3$ & 6$^\circ$ & 58.6; 29.4; 12.0 & 6.43 & 3.9 &  $2.4 \times 10^{22}$ & $ t \bar{t}$  & $10^3$   & 6$^\circ$  & 53.4; 32.5; 14.1  & 6.60 & 4.0   & $3.6 \times 10^{21}$ \\
\hline
\end{tabular}
\end{center}
\vspace{-0.2cm} \caption{The relevant parameter summary to calculate
the Super-Kamiokande constraints on $\Gamma_A$ for different DM
annihilation channels and masses. The units of $m_D$ and $\phi_\mu$
are GeV and $10^{-15} {\rm cm^{-2} sec^{-1}}$. } \label{N90phi}
\end{table}

The neutrino induced upgoing muon events in the Super-Kamiokande can
be divided into three categories: stopping, non-showering
through-going  and showering through-going \cite{SK}. The fraction
of each upgoing muon category as a function of parent neutrino
energy $E_{\nu_{\mu}}$ has been shown in Fig. 2 of Ref. \cite{SK}.
Then we use  $d N^\alpha_{\nu_\mu}/d E_{\nu_\mu}$ to calculate the
fraction of each category $F^i$ as listed in Table \ref{N90phi}.
Once $F^i$ is obtained, the  90\% confidence level (CL) upper
Poissonian limit $N_{90}$ can be derived through the following
formulas \cite{SK}:
\begin{eqnarray}
90 \%  = \frac{\int_{\nu_s = 0}^{N_{90}} L(n_{obs}^{i}|\nu_s) d
\nu_s}{\int_{\nu_s = 0}^{\infty} L(n_{obs}^{i}|\nu_s) d \nu_s}
\label{N90}
\end{eqnarray}
and
\begin{eqnarray}
L(n_{obs}^{i}|\nu_s) = \prod_{i=1}^3 \frac{(\nu_s F^i +
n_{BG}^i)^{n_{obs}^{i}}}{n_{obs}^{i}!} e^{-(\nu_s F^i + n_{BG}^i)}
\;,
\label{L}
\end{eqnarray}
where $\nu_s$ is the expected real signal. The number of observed
events of each category $n_{obs}^{i}$ and the expected background of
each category $n_{BG}^i$ for different DM masses and cone
half-angles can be found in Tables 1 and 2 of Ref. \cite{SK}. With
the help of Eqs. (\ref{N90}) and (\ref{L}), we estimate the 90\% CL
upper Poissonian limit on the number of upgoing muon events $N_{90}$
and the corresponding 90\% CL upper Poissonian limit of upgoing muon
flux $\phi_\mu = N_{90}/(1.67 \times 10^{15} {\rm cm^2 sec})$ as
shown in Table \ref{N90phi}.

\begin{figure}[htb]
\begin{center}
\includegraphics[scale=0.7]{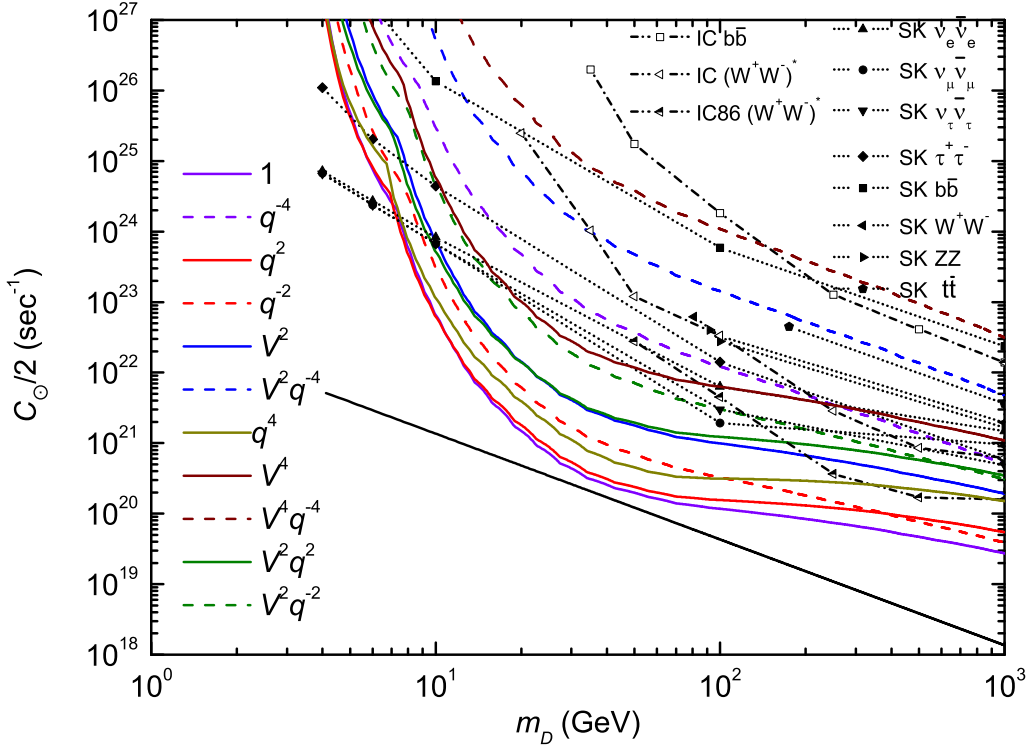}
\end{center}
\vspace{-0.5cm} \caption{The current Super-Kamiokande and IceCube
constraints on $C_\odot/2$ with the assumption $\Gamma_A =
C_\odot/2$ and the predicted maximal DM solar capture rates
$C_\odot/2$ from Fig. \ref{CsunHL} for different $F_{\rm
DM}^2(q,v)$. The black solid line describes the equilibrium
condition for $\langle \sigma v \rangle \approx 3.0\times 10^{-26}
{\rm cm}^3 {\rm sec}^{-1}$.} \label{Csun}
\end{figure}

With the help of Eq. (26) in Ref. \cite{Guo}, we numerically
calculate the neutrino induced muon flux from per DM pair
annihilation in the Sun. Then we directly derive the
Super-Kamiokande constraints on $\Gamma_A$ from the $\phi_\mu$
values as listed in Table \ref{N90phi}. In Fig. \ref{Csun}, we plot
these results with the dotted lines and the predicted maximal DM
solar capture rates $C_\odot/2$ from Fig. \ref{CsunHL} for different
$F_{\rm DM}^2(q,v)$. It should be mentioned that $\Gamma_A =
C_\odot/2$ has been assumed in Fig. \ref{Csun}. As shown in Eq.
(\ref{GammaA}), the assumption $\Gamma_A = C_\odot/2$ holds if
$t_\odot \sqrt{ C_\odot C_A} \gg 1$. For the usual $s$-wave
thermally averaged annihilation cross section $\langle \sigma v
\rangle \approx 3.0\times 10^{-26} {\rm cm}^3 {\rm sec}^{-1}$
deduced from the DM relic density, we find that $t_\odot \sqrt{
C_\odot C_A} \geq 3.0$ (namely $\tanh^2[t_\odot \sqrt{ C_\odot C_A}]
\geq 0.99$) requires $C_\odot/2 \geq 4.3 \times 10^{22}/(m_D/1{\rm
GeV})^{3/2} {\rm sec}^{-1}$ which has been plotted in Fig.
\ref{Csun} with the black solid line. Therefore the predicted
$C_\odot/2$ above this line in Fig. \ref{Csun} will satisfy the
assumption $\Gamma_A = C_\odot/2$ when $\langle \sigma v \rangle
\gtrsim 3.0\times 10^{-26} {\rm cm}^3 {\rm sec}^{-1}$. In addition
to the Super-Kamiokande experiment, the IceCube collaboration has
also reported the upper limits on the DM annihilation rate
$\Gamma_A$ for the $\bar{b} b$ and $W^+W^-$ ($\tau^+ \tau^-$ below
$m_D = 80.4$ GeV) channels in Table I of Ref. \cite{IC1}. We plot
these results with the dash-dotted lines in Fig. \ref{Csun}. The
IC86$(W^+W^-)^*$ line shows the expected 180 days sensitivity of the
completed IceCube detector \cite{IC2}. Recently, the ANTARES
neutrino telescope \cite{ANTARES} has reported the first results
which are comparable with those obtained by the Super-Kamiokande
\cite{SK} and IceCube \cite{IC1,IC2}. It is shown that the upper
limits on $C_\odot$ ($\sigma_n$) from the Super-Kamiokande and
IceCube are weaker than those from the current direct detection
experiments for the usual SI DM-nucleus interaction. However, our
numerical results in Fig. \ref{Csun} clearly show the
Super-Kamiokande and IceCube may give more stringent constraints
than the XENON100 experiment for several momentum and velocity
dependent DM-nucleus interactions with $m_D \gtrsim 10$ GeV and the
assumption $\Gamma_A = C_\odot/2$.

In Fig. \ref{Csun}, one may find that the Super-Kamiokande
experiment can significantly constrain the low mass DM for all DM
form factors in Table \ref{FDM} when the DM particles dominantly
annihilate into neutrino pairs or $\tau^+ \tau^-$. If $m_D \gtrsim
20$ GeV, both Super-Kamiokande and IceCube can not constrain any
momentum and velocity dependent case except for the $V^4q^{-4}$
case, when the annihilation channel is the $b \bar{b}$, and the $1,
q^2, q^{-2}, q^4, V^2, V^2q^2$ and $V^2q^{-2}$ cases for any
annihilation channel. The $V^4q^{-4}$ and $V^2q^{-4}$ cases can be
significantly constrained by the above two experiments if the DM
annihilation final states are neutrinos, tau leptons or gauge
bosons. For the $W^+W^-$ channel, the IceCube gives the stronger
constraint than the Super-Kamiokande when $m_D \gtrsim 100$ GeV. The
future IceCube result IC86$(W^+W^-)^*$ has ability to constrain the
$q^{-4}, V^2, V^4, V^2q^2$ and $V^2q^{-2}$ cases with $m_D \gtrsim
200$ GeV. Since $C_\odot$ is proportional to $\sigma_n$, the upper
limits on $C_\odot$ in Fig. \ref{Csun} will move downward if
$\sigma_n$ in Fig. \ref{DI} becomes smaller. The Super-Kamiokande
experiment can still constrain the momentum and velocity dependent
DM-nucleus interactions for the low DM mass region even if
$\sigma_n$ is reduced by 2 orders.

\section{The mediator mass $m_\phi$ effect on $\sigma_n$ and $C_\odot$ }

In Sec. \ref{EO}, we have taken two extreme scenarios for the
mediator mass $m_\phi$: $m^2_\phi \gg q^2$ and $m^2_\phi \ll q^2$.
Here we shall consider the $m^2_\phi \sim {\cal{O}}(q^2)$ scenario
and discuss the $m_\phi$ effect on $\sigma_n$ and $C_\odot$. In this
case, the momentum and velocity dependent DM form factors are
relevant to $m_\phi$. It is found that the $m_\phi$ dependent DM
form factors $F_{\rm DM}^2(q,v,m_\phi)$ can be written by the
product of the third column of Table \ref{FDM} and a factor $(q_{\rm
ref}^2 + m_\phi^2)^2/(q^2 + m_\phi^2)^{2}$. The two DM form factors
in each row of Table \ref{FDM} are two extreme cases of the $m_\phi$
dependent DM form factor $F_{\rm DM}^2(q,v,m_\phi)$. For example,
one can easily obtain $F_{\rm DM}^2(q,v) = q^2/q_{\rm ref}^2$ with
$m^2_\phi \gg q^2,q^2_{\rm ref}$ and $F_{\rm DM}^2(q,v) = q_{\rm
ref}^2/q^2$ with $m^2_\phi \ll q^2,q^2_{\rm ref}$  from $F^2_{\rm
DM} (q,v,m_\phi) = (q^2/q_{\rm ref}^2) (q_{\rm ref}^2 +
m_\phi^2)^2/(q^2 + m_\phi^2)^{2}$. Therefore, we have 6 kinds of
$m_\phi$ dependent DM form factors $F^2_{\rm DM} (q,v,m_\phi)$. Here
we use $1m_\phi, q^2m_\phi, V^2m_\phi, q^4m_\phi, V^4m_\phi$ and
$V^2q^2m_\phi$ to express them.

\begin{figure}[htb]
\begin{center}
\includegraphics[scale=0.45]{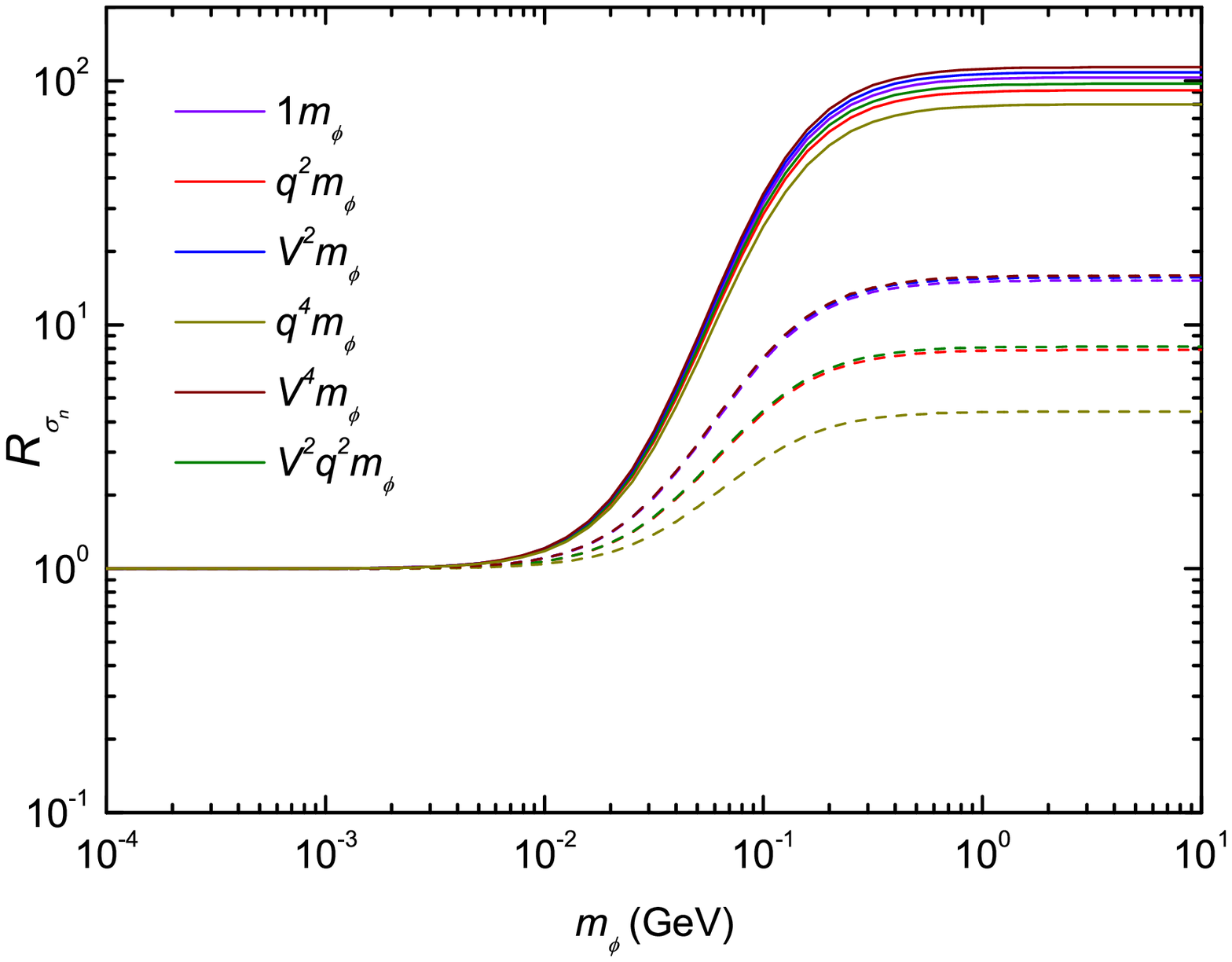}
\includegraphics[scale=0.45]{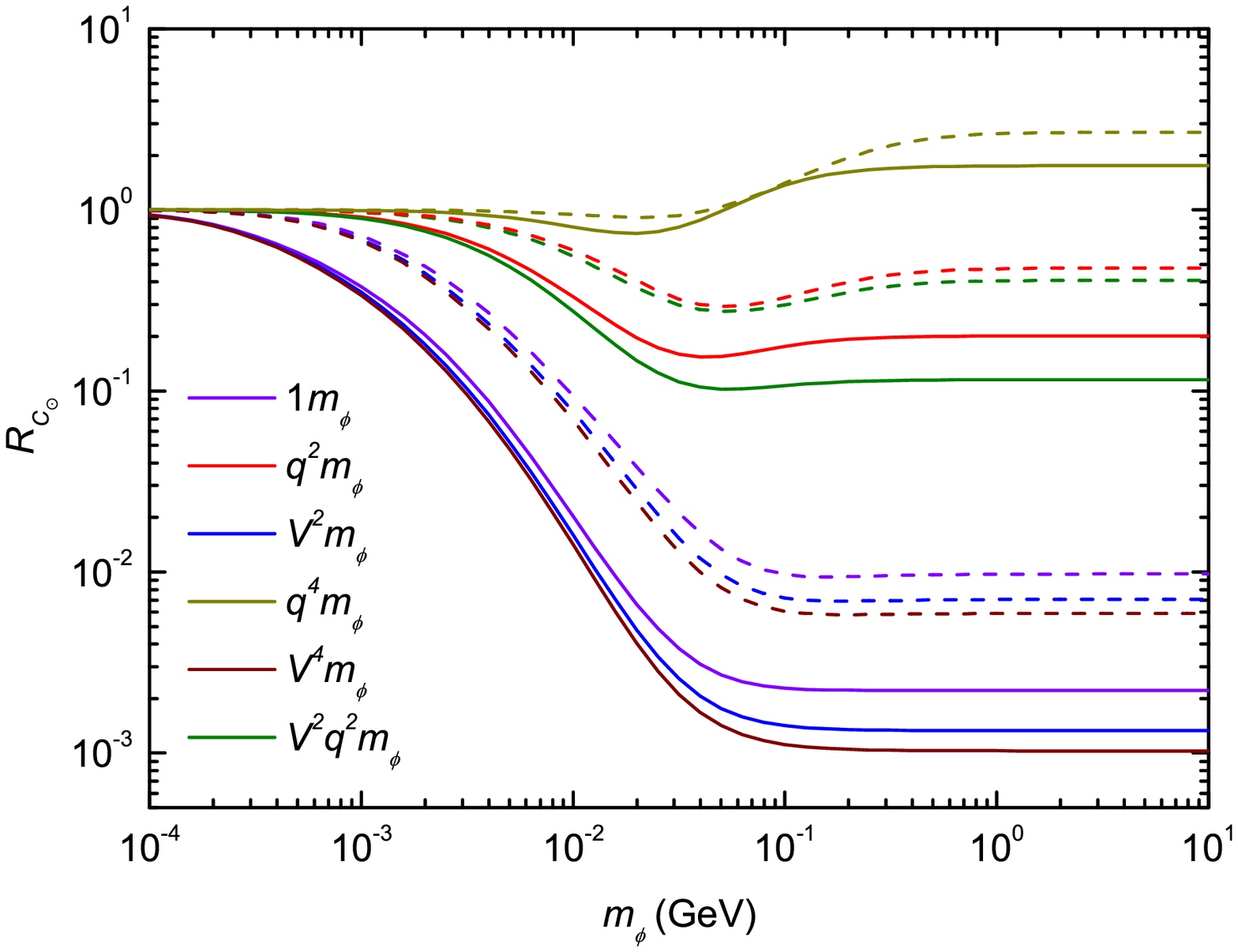}
\end{center}
\vspace{-0.5cm}\caption{ The predicted $\sigma_n$ and $C_\odot$ as a
function of $m_\phi$ for $m_D = 10$ GeV and $m_D = 100$ GeV. The
parameters $R_{\sigma_n}$ and $R_{C_\odot}$ are defined as
$R_{\sigma_n} \equiv \sigma_n/\sigma_n(m_\phi^2 \ll q^2,q^2_{\rm
ref})$ and $R_{C_\odot} \equiv C_\odot/ C_\odot(m^2_\phi \ll
q^2,q^2_{\rm ref})$. The solid and dashed lines describe $m_D = 10$
GeV and  $m_D = 100$ GeV cases, respectively.} \label{R1}
\end{figure}

Using the above 6 $m_\phi$ dependent DM form factors, we calculate
$\sigma_n$ and $C_\odot$ for two representative DM masses: $m_D =
10$ GeV and $m_D = 100$ GeV. Our numerical results have been shown
in Fig. \ref{R1}. The parameters $R_{\sigma_n}$ and $R_{C_\odot}$ in
Fig. \ref{R1} are defined as $R_{\sigma_n} \equiv
\sigma_n/\sigma_n(m_\phi^2 \ll q^2,q^2_{\rm ref})$ and  $R_{C_\odot}
\equiv C_\odot/C_\odot(m^2_\phi \ll q^2,q^2_{\rm ref})$.
$\sigma_n(m^2_\phi \ll q^2,q^2_{\rm ref})$ and $C_\odot(m^2_\phi \ll
q^2,q^2_{\rm ref})$ denote the DM scattering cross section and solar
capture rate in the $m^2_\phi \ll q^2,q^2_{\rm ref}$ case,
respectively. One may see from  Fig. \ref{R1} (left panel) that
$\sigma_n$ will remarkably increase as $m_\phi$ increases when
$m_\phi \sim q_{\rm ref} = 0.1$ GeV. For $m_\phi \lesssim 0.01$ GeV
and $m_\phi \gtrsim 0.2$ GeV, the predicted $\sigma_n$ is
insensitive to $m_\phi$. These features can be easily understood
from the forms of $F_{\rm DM}^2(q,v,m_\phi)$. For the light DM mass
$m_D = 10$ GeV, our numerical results show that 6 kinds of $m_\phi$
dependent DM form factors can produce the similar curves. As shown
in Fig. \ref{R1} (right panel), the predicted $C_\odot$ approaches
to a constant as well as $\sigma_n$ if $m_\phi \gtrsim 0.2$ GeV. For
$m_\phi < 0.2$ GeV, $C_\odot$ can usually decrease as $m_\phi$
increases. We find that the $q^{2}m_\phi$, $q^4m_\phi$ and $V^2
q^{2}m_\phi$ cases have the minimums around $m_\phi \thickapprox
0.04$ GeV for $C_\odot$. In fact, the DM solar capture rates with a
fixed $\sigma_n$ in the $q^{2}m_\phi$, $q^4m_\phi$ and $V^2
q^{2}m_\phi$ cases are the monotone decreasing functions of
$m_\phi$. Therefore the minimums arise from the monotone increasing
$\sigma_n$. When the $\sigma_n$ increase is larger than the
$C_\odot$ (with a fixed $\sigma_n$) decrease, we can derive
$R_{C_\odot} >1$, just like the $q^4m_\phi$ case in the right panel
of Fig. \ref{R1}. In terms of the results in Fig. \ref{R1}, the DM
solar capture rate in the $m_\phi$ dependent scenario will quickly
move from the dashed line to the corresponding color solid line as
$m_\phi$ increases in Fig. \ref{Csun}. When $m_\phi \gtrsim 0.2$
GeV, the $m_\phi$ dependent scenario will approach to the heavy
mediator mass scenario.

\section{Discussions and Conclusions}

So far, we have used the usual Helm nuclear form factor for
$F_N^2(q)$ in Eqs. (\ref{R}) and (\ref{sigmaA}) to calculate
$\sigma_n$ and $C_\odot$. In fact, the exact $F_N^2(q)$ contains the
standard SI nuclear form factor (Helm nuclear form factor) and an
important correction from the angular-momentum of unpaired nucleons
within the nucleus for the NR operators ${\cal O}_3$ and ${\cal
O}_4$ in Eq. (\ref{operator}) \cite{Fitzpatrick:2012ix}. The
correction is comparable with the standard SI nuclear form factor
for nuclei with unpaired protons and neutrons when $m_D \gtrsim
m_N$. By use of the relevant formulas in Appendix A of Ref.
\cite{Fitzpatrick:2012ix}, we numerically calculate this correction
contribution to the XENON100 and XENON10 experiments and find that
it is smaller than 10\% and can be neglected for our analysis about
$\sigma_n$. In the previous sections, the predicted $C_\odot$ arises
from the contributions of $^1$H, $^4$He, $^{12}$C, $^{14}$N,
$^{16}$O, Ne, Mg, Si, S and Fe. Since these elements or dominant
isotopes have not the unpaired protons and neutrons within the
nucleus, our numerical results about $C_\odot$ are not significantly
changed.

In conclusion, we have investigated the SI momentum and velocity
dependent DM-nucleus interactions and discussed their effects on
$\sigma_n$ and $C_\odot$. In terms of the NR effective theory, we
phenomenologically discuss 10 kinds of momentum and velocity
dependent DM form factors $F_{\rm DM}^2(q,v)$. Using these DM form
factors, we have recalculated the corresponding upper limits on
$\sigma_n$ from the XENON100 and XENON10 experimental results. Each
upper limit on $\sigma_n$ can be used to calculate the corresponding
maximal DM solar capture rate $C_\odot$. Our numerical results have
shown that the momentum and velocity dependent DM form factor cases
can give larger DM annihilation signals than the usual contact
interaction case for almost the whole parameter space. The light
mediator mass scenario can usually produce the larger $C_\odot$ than
the corresponding heavy mediator mass scenario except for the $q^4$
and 1 cases. On the other hand, we have also deduced the
Super-Kamiokande's constraints on $C_\odot/2$ for 8 typical DM
annihilation channels with the equilibrium assumption $\Gamma_A=
C_\odot/2$. In contrast to the usual contact interaction, the
Super-Kamiokande and IceCube experiments can give more stringent
limits on $\sigma_n$ than the latest XENON100 experiment for several
momentum and velocity dependent DM form factors when $m_D \gtrsim
10$ GeV and $\Gamma_A = C_\odot/2$. In addition, we have also
considered 6 kinds of $m_\phi$ dependent DM form factors and
analyzed their effects on $\sigma_n$ and $C_\odot$. We find that
$C_\odot$ will quickly move from the light mediator mass scenario to
the corresponding heavy mediator mass scenario as $m_\phi$
increases. When $m_\phi \gtrsim 0.2$ GeV, the $m_\phi$ dependent
scenario will approach to the corresponding heavy mediator mass
scenario.

\acknowledgments

This work is supported in part by the National Basic Research
Program of China (973 Program) under Grants No. 2010CB833000; the
National Nature Science Foundation of China (NSFC) under Grants No.
10821504 and No. 10905084; and the Project of Knowledge Innovation
Program (PKIP) of the Chinese Academy of Science.

\end{document}